\documentclass[fleqn,10pt]{wlscirep}
\usepackage[utf8]{inputenc}
\usepackage[T1]{fontenc}
\usepackage{subcaption}

\usepackage{xcolor}
\newcommand {\out}[1]{}

\title{Smartphone-based syndromic surveillance against COVID-19 clusters in Japan}

\author[1]{Shohei Hisada}
\author[1]{Taichi Murayama}
\author[2]{Kota Tsubouchi}
\author[2]{Sumio Fujita}
\author[1]{Shuntaro Yada}
\author[1]{Shoko Wakamiya}
\author[1,*]{Eiji Aramaki}

\affil[1]{Nara Institute of Science and Technology (NAIST), Japan}
\affil[2]{Yahoo Japan Corporation, Japan}

\affil[*]{aramaki@is.naist.jp}

\begin{abstract}
Two clusters of the coronavirus disease 2019 (COVID-19) were confirmed in Hokkaido, Japan, in February 2020. To capture these clusters, this study employed web search query logs and user location information from smartphones. Initially, we anonymously identified the smartphone users who used a web search engine (i.e., Yahoo! JAPAN Search) to search for COVID-19 or its symptoms. We regarded these searchers as web searchers who were suspicious of their own COVID-19 infection (WSSCI). Subsequently, we extracted the location of WSSCI via the smartphone application and compared the spatio-temporal distribution of WSSCI with the actual location of the two known clusters. In the early stage of cluster development, we can confirm several WSSCI. However, our approach was accurate only in that stage and became biased after the public announcement of the cluster development. 
When other cluster-related resources, such as detailed population statistics, are not available, the proposed metric can capture hints of emerging clusters.
\end{abstract}

\begin{document}

\flushbottom
\maketitle
\thispagestyle{empty}

\noindent 

\section*{Introduction}


In 2019 and 2020, a disease caused by a novel coronavirus called SARS-CoV-2 has spread worldwide~\cite{WHO2019}.
To control the rapid spreading of this coronavirus disease 2019 (COVID-19), pandemic management tasks, including optimizing arrangements of medical service supply, health, and medical information dissemination and control and development of relevant laws and rules, are important for public health authorities and relevant governmental organizations.
Among various tasks, the cluster response, which intends to detect small groups of infected people in a large community, is significant in the early stage of a pandemic, because this enables health authorities to restrict the spread of the virus~\cite{MHLW2019}.

On February 25, 2020, the Ministry of Health, Labour and Welfare (MHLW), Japan, convened a team of about 30 specialists to identify clusters; they identified 13 clusters until March 17, 2020~\cite{ClusterMap2}. 
There were possibly more clusters because young people who tend to have mild symptoms, compared to the old ones~\cite{who20204}, actively move around to work or go to school.
Considering the business and social activities of young people, it is crucial to capture the slight signals of the infected youth. 

This context motivated us to leverage the usage logs of smartphones, through which we could collect information from anyone who used a location-aware smartphone application, anytime and anywhere.
We employed two types of smartphones usage logs: web search query logs and location information. 
We also deployed web search query logs from PCs, as described in the ``Material and Methods'' section, to encompass people who searched for the disease on their PCs at home.
First, we assumed that a person who might have COVID-19 attempts to obtain detailed information on the disease and its symptoms through a web search. 
Considering that, we designed 63 query patterns, such as ``likely to be corona,'' ``cough'' and ``corona,'' ``diarrhea'' and ``new type,'' and ``coughing up phlegm'' and ``new type pneumonia.''
We defined web users whose queries were related to the symptoms of COVID-19 as \textbf{web searchers who were suspicious of their own COVID-19 infection} (\textbf{WSSCI}).
Second, we prospectively extracted location information of WSSCI from the ``Yahoo! JAPAN App~\cite{yahooapp},'' which is one of the most popular smartphone applications in Japan and hosts many services, such as web search and weather report. 
Note that the location data were collected from users who approved the research purpose uses of their location information.
Subsequently, we counted the number of WSSCI in each day and each area based on their location information. 

A previous study~\cite{Ginsberg2009} demonstrated that symptom-related search queries have an advantage to capture the early signals of infectious diseases. In addition, some studies~\cite{Lampos2015,Zhang2019,Ning2019} attempted to utilize search queries for nowcasting or predicting influenza epidemic. 
A recent study~\cite{Samaras2020} reports that both search queries and social network data can produce precise and usable estimation of the influenza development by investigating which kind of data source leads to better results. 
A recent work-in-progress paper also uses web search queries to predict country-level COVID-19 epidemic~\cite{lampos2020tracking}, and it applies a prediction model based on search query trends within countries that observed COVID-19 epidemics in other countries, without having experienced any of them yet.
We, in contrast, utilized web search query logs \textit{per user} to detect WSSCI and aggregate their location histories to identify locations they visit or pass, resulting in COVID-19 \textit{cluster} detection.

We believe that our approach is suitable for COVID-19 cluster detection because smartphone applications are widely used nowadays. 
This study investigates the feasibility of our approach through case studies of the COVID-19 clusters occurred in February 2020, in Hokkaido, Japan. 
The COVID-19 pandemic indeed made us realize that obtaining reliable information of the current status during a pandemic crisis is challenging. 
However, even in such a low resource condition, smartphone users are still available and can be regarded as a type of \textit{social sensors} who voluntarily report current events in real time whether they realize it or not. 
Therefore, to take advantage of social sensors, it is essential to examine the validity of the WSSCI-based approach in advance. 

\section*{Results}
\begin{figure}[t]
    \centering
    \includegraphics[clip,width=0.9\linewidth]{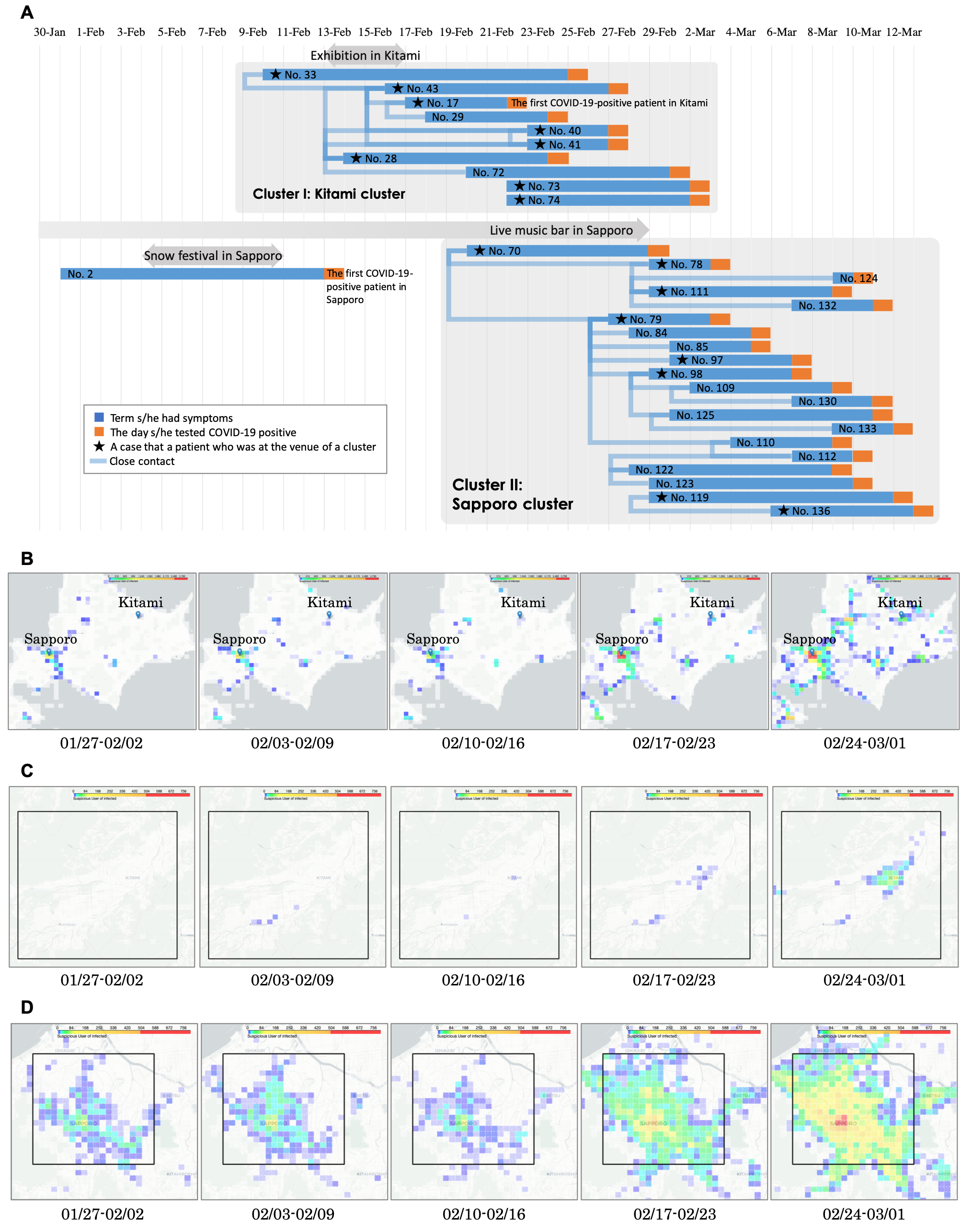}
    \caption{\textbf{Occurrence status of the two clusters in Hokkaido and spatio-temporal distribution of the number of WSSCI per half grid (WSSCIphg) in Hokkaido.} (A): Timeline of outbreaks of two clusters in Hokkaido, Cluster I: Kitami cluster and Cluster II: Sapporo cluster, based on the Hokkaido government official announcement. 
    The existence of Kitami cluster and Sapporo cluster was admitted on March 9 and March 18, 2020, respectively. (B): Total weekly WSSCIphg throughout every 400 half grid squares in Hokkaido, Japan. (C-D): Total weekly WSSCIphg throughout every four half grid squares in Kitami and Sapporo, respectively, surrounded by a rectangle. In (B-D), the colors of grid squares represent the total weekly WSSCIphg throughout every several half grid squares, of which red is large and blue is small. The blue marker denotes the location of a cluster.}
    \label{fig1}
\end{figure}

The location information of WSSCI were extracted from a smartphone application (mentioned in the ``Material and Methods'' section).
We focused on the area of Hokkaido where two COVID-19 clusters had been reported in March 2020, according to cluster maps~\cite{ClusterMap1,ClusterMap2} released by the MHLW. 
Figure~\ref{fig1}A shows the timelines of the clusters based on official announcements by the Hokkaido government~\cite{hokkaido-rep}, Kitami cluster~\cite{kitami-rep} and Sapporo cluster~\cite{sapporo-rep}.
Figures~\ref{fig1}B, ~\ref{fig1}C, and ~\ref{fig1}D show the spatio-temporal distributions of WSSCI in the areas of Hokkaido, Kitami, and Sapporo, respectively.
In Figure~\ref{fig1}B, WSSCI hot spots had existed in the two cluster areas before they were reported, suggesting that WSSCI could be a clue of the cluster arrival.
However, there is no cluster in several WSSCI hot spots near the central and the southern part of Hokkaido.
This suggests that not all hot spots are clusters, exposing the limitation of our approach.

\begin{figure}[t]
    \centering
    \includegraphics[clip,width=0.9\linewidth]{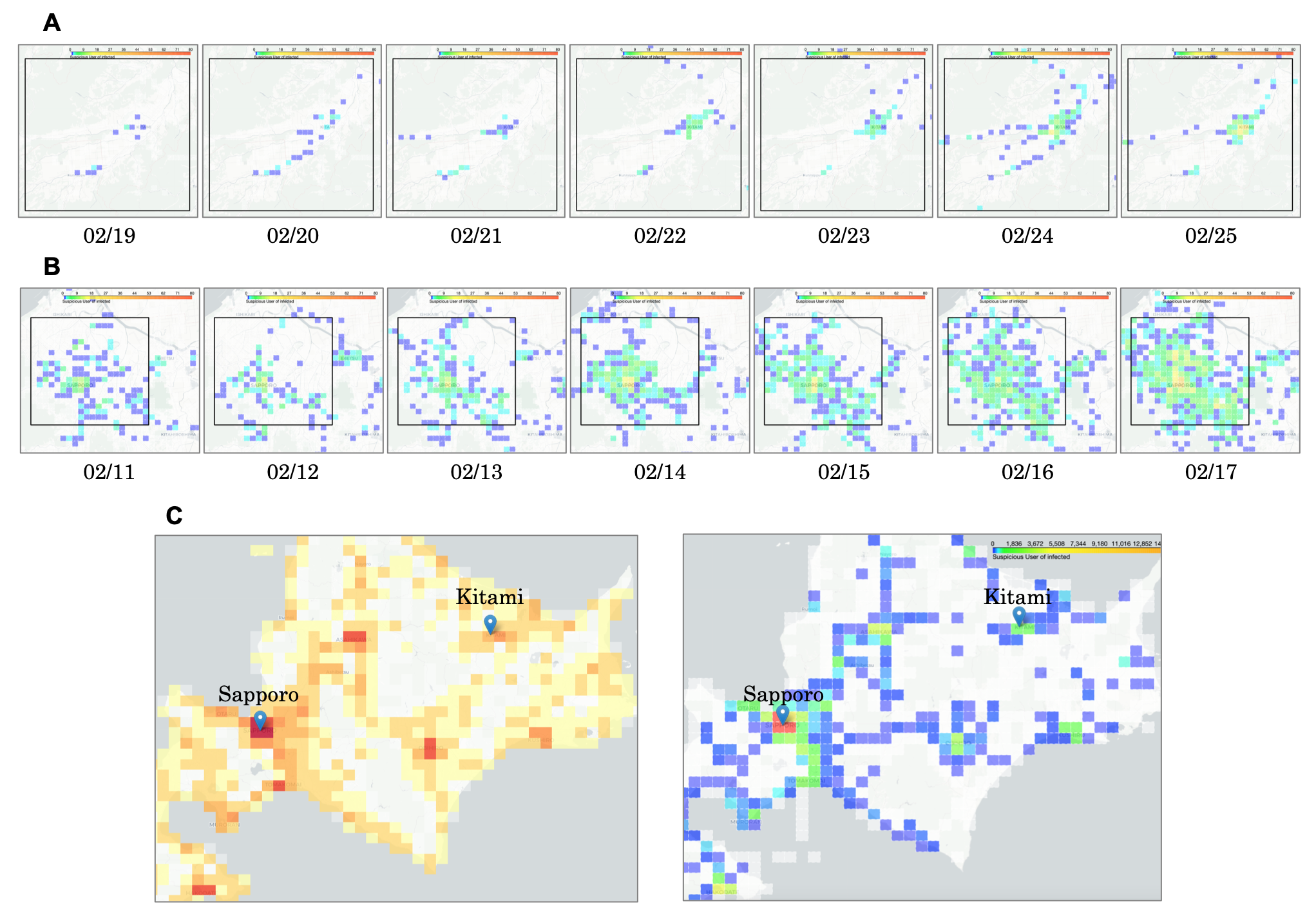}
    \caption{\textbf{Spatio-temporal distribution of the number of WSSCI per half grid (WSSCIphg), population distribution and total WSSCIphg in Hokkaido.} (A-B): The distribution of the number of WSSCIphg throughout every four half grid squares between 02/19 and 02/25 in Kitami surrounded by a rectangle and between 02/11 and 02/17 in Sapporo surrounded by a rectangle, respectively. 
    The figure shows 3 days before and 3 days after the announcement of the first patient positive in each city (a total of 7 days).
    (C) Population distribution in Hokkaido (left) and total WSSCIphg throughout every 400 half grid squares between January 27 and March 1, 2020 (right).}
    \label{fig2}
\end{figure}

\out{
\begin{figure}[t]
    \begin{minipage}[t]{1.0\hsize}
        \centering
        \includegraphics[clip,width=0.99\linewidth]{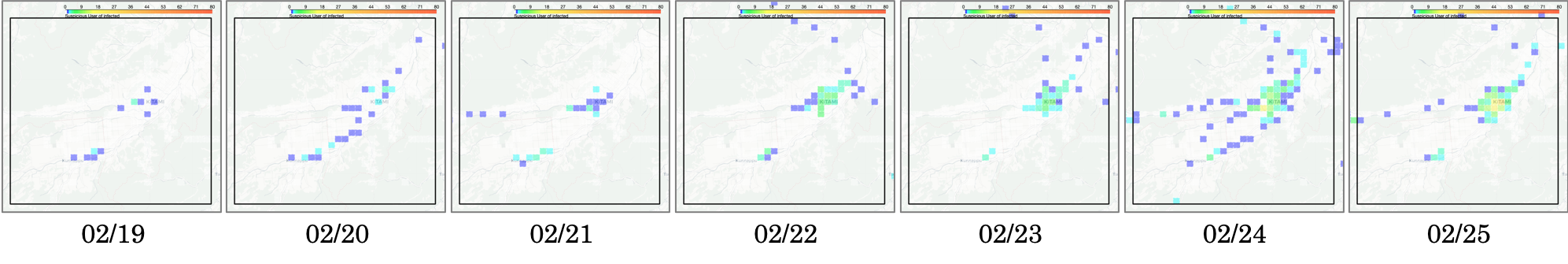}
        \subcaption{Between 02/19 and 02/25 in Kitami surrounded by a rectangle.}
        \label{fig2kitami}
    \end{minipage}\\ 
    \begin{minipage}[t]{1.0\hsize}
        \centering
        \includegraphics[clip,width=0.99\linewidth]{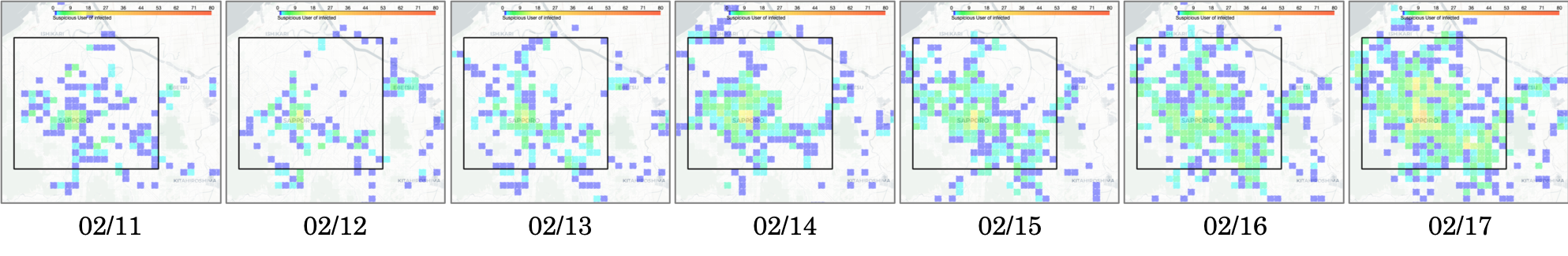}
        \subcaption{Between 02/11 and 02/17 in Sapporo surrounded by a rectangle.}
        \label{fig2sapporo2}
    \end{minipage}
    \begin{minipage}[t]{1.0\hsize}
    \centering
        \includegraphics[clip,width=0.6\linewidth]{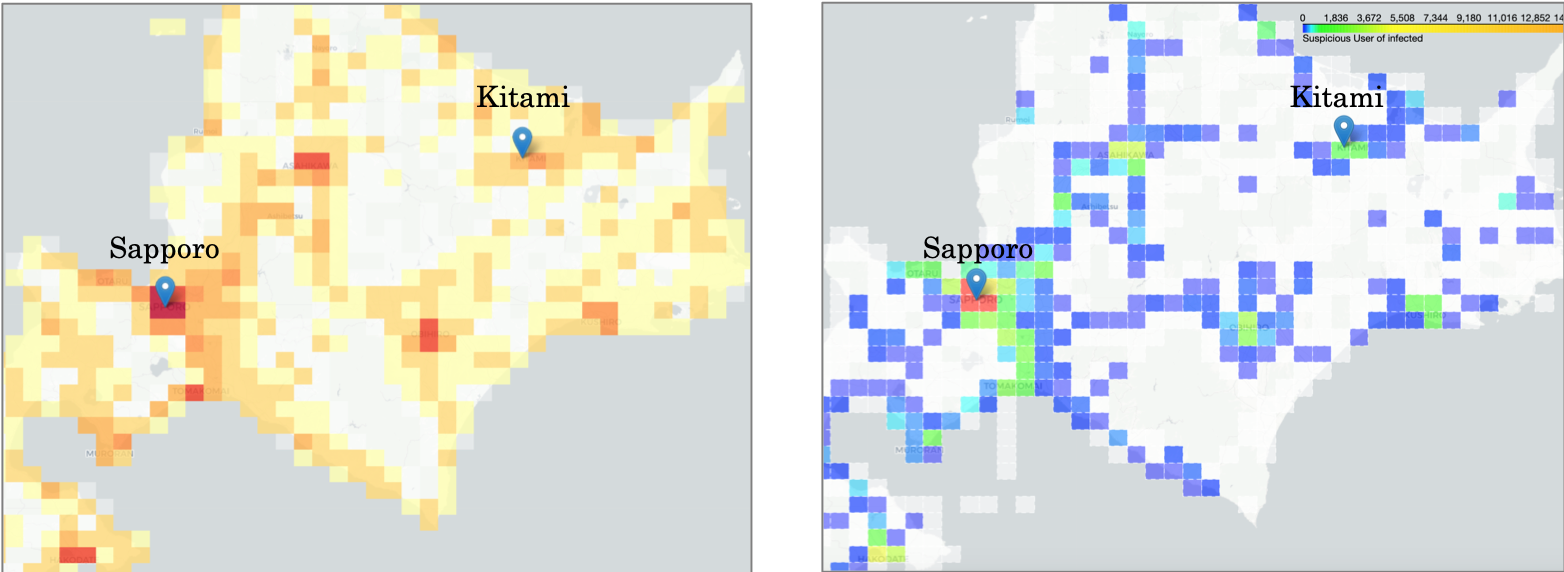}
    \subcaption{Population distribution (left) and all WSSCI distribution (right)}
        \label{fig3}
    \end{minipage}
    \caption{(a) (b) Spatio-temporal distribution of the number of WSSCI per half grid (WSSCIphg) in Hokkaido, 
    WSSCIphg throughout every four half grid squares in the two cities, Kitami and Sapporo. 
    The figure shows 3 days before and 3 days after the announcement of the first patient positive in each city (a total of 7 days).
    (c) Population distribution in Hokkaido (left) and total WSSCIphg throughout every 400 half grid squares between January 27 and March 1, 2020 (right).}
    \label{fig2}
\end{figure}
}

\subsection*{Cluster I: Kitami cluster}

The first cluster comprises more than 10 cases related to the participants of an exhibition and a dinner party in Kitami between February 13 and 15, 2020. Eight participants have been confirmed to be infected in this event, as shown in the upper part of Figure~\ref{fig1}A. 
Starting with the case reported in Kitami on February 22 (No. 17), five participants (Nos. 17, 33, 43, 73, and 74) including the first case in Kitami and three participants from Sapporo (Nos. 28, 40, and 41) have been confirmed to be infected. Regarding the interpersonal relations, person No. 29 was in close contact with person No. 17, and person No. 72 with person No. 28.

Before the breaking news of the first patient infected with the coronavirus in Kitami City on February 22, 2020, we could confirm several WSSCI between February 3 and 16, 2020, in Figure~\ref{fig1}C, which were not confirmed at all in the previous week, January 27 to February 2, 2020.
After the breaking news of February 22, the WSSCI spread across this area. 
From this fact, it can be inferred that the appearance of WSSCI might indicate the cluster existence, demonstrating the feasibility of the WSSCI-based cluster prediction. 
Figure~\ref{fig2}A shows the spatio-temporal distribution of WSSCI; we can see some hot spots of WSSCI on February 19, 2020.
Considering the initial status of this area between January 27 and February 2, 2020, that is the nonexistence of WSSCI, as shown in Figure~\ref{fig1}C, these hot spots might provide us hints about the cluster existence or the possibility of cluster occurrence in the near future.

\subsection*{Cluster II: Sapporo cluster}

In another cluster in Hokkaido, as shown in the lower part of Figure~\ref{fig1}A, eight people have been confirmed to be infected in a live music bar in Sapporo. Starting with a store clerk, No. 70, who was tested positive on February 29, 2020, after visiting a medical office, other two clerks (Nos. 78 and 79) were also positive on March 3, 2020.
The bar remained open until February 29, 2020, when the first person tested positive. Two guests (Nos. 119 and 136) who visited the bar on February 25, 2020, one from Otaru and the other from Kitami. The other three guests (Nos. 97, 98, and 111) who visited the bar on February 26, 2020, have been confirmed to be infected also.
 In addition, the infection spread to a total of 12 close contacts of the people who had been in the live music bar.
 
In Sapporo, the cluster occurrence was first reported on March 17, 2020~\cite{ClusterMap2}. 
Unlike the Kitami cluster, the first patient who was tested positive in Sapporo was not directly related to the Sapporo cluster. 
Furthermore, there had already been 13 patients tested positive in Sapporo and 66 patients tested positive in the entire Hokkaido before February 29, 2020, when the first case of the Sapporo cluster was confirmed. 
Therefore, although we can confirm that WSSCI clearly increased and spread across Sapporo during the week when the first positive case of Sapporo cluster (February 24 to March 1, 2020) was confirmed, many WSSCI had already been spreading across the area before the week of February 10 to 16, as shown in Figure~\ref{fig1}D. 

To investigate more detailed trends of WSSCI, Figure~\ref{fig2}B shows the daily spatial distribution of WSSCI in Sapporo from February 11 to February 17, 2020. 
On February 11, 2020, several WSSCI had already existed, indicating this area is a part of densely populated city in terms of both population statistics and WSSCI, as shown in Figure~\ref{fig2}C .
On February 14, 2020, the occurrence of the first case of a suspected patient in this area made national headlines. 
However, there had not been a cluster at that time yet (unlike Cluster I: Kitami cluster). 
In such a situation, many WSSCI had appeared after the report of the suspected patient on February 14, 2020.
The reason of this sudden burst of WSSCI is that many people attempted to obtain more information about COVID-19 even they did not have any symptoms. 
This indicates that the WSSCI-based approach could be easily biased by varying media coverage.

\section*{Discussion}

This study attempted to detect early clusters by identifying people who suspected their own COVID-19 infection using web search query logs and by extracting their location information.
To identify possible COVID-19-positive patients, using web search query logs is one of the advantages of the WSSCI-based approach. 
Although statistics of inpatients and outpatients have been reported, COVID-19 patients with mild symptoms who do not go to the hospital are difficult to identify. 
In fact, even patients with mild symptoms can transmit the virus to others; it is therefore highly required to find these patients. Fortunately, our approach can consider such patients through their location information.

Moreover, to use location information of WSSCI is another advantage of our approach. 
Location information of all users can only reveal us crowded areas, which can also be done through other resources, such as population statistics. 
However, it is more important to know ``where people who suspect they might be infected with the disease'' are concentrated especially in the early stage of the spread of the infectious disease.
In addition to COVID-19, these data could help us control the spread of other infectious diseases or recurrent coronavirus in the future at an early stage. 
We believe that the proposed method could enhance these attempts. 

Another advantage of the WSSCI-based approach is the real time nature of the users.
In reality, people move from one city to another to avoid COVID-19. It might cause a dynamic change of population distribution, as shown in Figure~\ref{fig2}C. 
Even in such a situation, the WSSCI-based approach can grasp ongoing local migrations. 
Additionally, we realized that it might be effective to support a real-time cluster surveillance in developing countries, where we can safely assume at least the existence of web search traffic in comparison to other smartphone applications (e.g., location-sharing social network services).

This first study for ongoing pandemic crisis has the advantages described above, but there are limitations that should be considered to apply syndromic surveillance. 
One limitation is that an area with high concentration of WSSCI does not always indicate the existence of a cluster. 
In fact, among several hot spots in Figure~\ref{fig1}B, the number of clusters officially detected was only two, causing an overestimation of clusters.
However, as mentioned above, it would be helpful to narrow down potential areas that might be clusters in the early stage. 

Another limitation is a bias caused by obtained varying media coverage. 
Whenever the mass media report the location of a COVID-19-positive patient, many people who are in areas adjacent to the reported location submit COVID-19-related queries to acquire additional information. 
Therefore, media could change the nature of WSSCI, suggesting that our approach would be more effective to detect potential cluster areas before any national headlines.

\section*{Methods}

\subsection*{Ethics statement}
All participants provided written (or electronically displayed) informed consent before participating in this study and agreed to the terms of the Conditions of the Use of Services provided by Yahoo! Japan services when they disclosed their location.
This study did not require the participants to be involved in any physical and/or mental intervention.
Participants’ information was unlinkable, anonymized, and deidentified prior to analysis. 
This research did not obtain identifiable private information, meaning that it was exempt from Institutional Review Board approval according to the Ethical Guidelines for Research of the Japanese national government.

\subsection*{Material}
We utilized web search query logs and location information from a smartphone application provided by Yahoo Japan Corporation~\cite{yahoocorp}, a company providing a variety of web services, including web search, news, weather forecasts, and shopping, as signals for syndromic surveillance.
The number of active monthly users of its services is 62.7 million for smartphones and 
21.4 million for PCs from January to June 2019, 2020, which are the largest in Japan, 
in comparison with YouTube, 56.3 million for smartphones and 11.6 million for PCs, 
or Facebook, 50.8 million for smartphones and 4.7 million for PCs, 
according to Nielsen NetView/Mobile NetView Customer Feed~\cite{YJPD_2019}.

\subsection*{Infection suspecting searcher identification using web search query logs}
We leveraged the search logs on all devices from Yahoo! Japan Search~\cite{yahoosearch} to identify search users who were possibly affected by the disease and were looking for information about the characteristic features of its symptoms to check whether their symptoms corresponded to typical cases of the disease.

Queries submitted by search users were matched against 63 pre-defined patterns, consisting of three single-term patterns and 60 double-term patters.
Single-term patterns included Japanese short phrases literally meaning ``likely to be corona,'' ``likely to be corona-virus,'' or ``likely to be new type pneumonia.''
Double-term patterns included one of three main terms succeeded by a facet term out of 20 patterns.
The main terms were ``corona\%,'' ``new type,'' or ``new type pneumonia,'' where $\%$ denoted wild card matching, as used in the ``like'' operator of the SQL language.
The facet terms included 18 Japanese phrases meaning one of the symptoms, namely,
``cough,'' ``diarrhea,'' ``coughing up phlegm,'' ``slight fever,'' ``headache,'' 
``cold,'' ``fevered,'' ``no fever,'' ``without fever,'' ``high fever,'' 
``develop fever,'' ``runny nose,'' ``chills,'' ``throat,'' ``chest,'' ``phlegm,'' and ``feel tired'' or ``weariness'', or two phrases meaning ``designated hospitals'' 
(which means hospitals validated by the local health authorities as specialized for treatments of COVID-19-infected patients)
or ``advice'' (which was presumably intended for special consultation services in charge of advising those who suspected to be infected with the virus).
These terms in double patterns were AND-combined; therefore, their order was irrelevant. A query was matched against a pattern if and only if all terms were matched against any terms in user queries.
Therefore, for queries matched against one of the 63 aforementioned patterns, 
anonymized user IDs of searchers were stored only when they had already approved 
the use of their smartphone location information for research purposes.
These searchers were defined as web searchers who are suspicious of their own COVID-19 infection (WSSCI).

\subsection*{User location information extraction}
The ``Yahoo Japan App'' is a popular smartphone application in Japan that hosts many services, including web search, news, weather report, and shopping, provided by Yahoo Japan Corporation.
For a portion of the app users who explicitly approved the use of their smartphone location information for research purposes, the app stores the GPS location information to the server.

Among such users, we mapped the number of WSSCI using the symptom suspected queries described in the previous subsection into each half grid square (500 m x 500 m) according to the location information.
Concerning the privacy, we used neither the IDs of each searcher nor their exact location information, but only the number of searchers in each day in the areas defined by the half grid square code system~\cite{stats},
which we called the number of WSSCI per half grid (WSSCIphg).

The number of searchers was counted only when the location information of the searchers was stored during their stay in the grid area.
We used three time spans in a day, namely, day time, 8:00 AM–3:59 PM, evening time, 4:00 PM–11:59 PM, and night time, 0:00 AM –7:59 AM,
in addition to the whole day time spans (0:00 AM–11:59 PM).





\section*{Data availability}
The data that support the findings of this study are available from the corresponding author upon reasonable request.

\section*{Code availability}
The custom code in these analyses are available from the corresponding author upon reasonable request.

\section*{Role of the funding source}
This study received no funding. The corresponding authors had full access to all data in the study and had final responsibility for the decision to submit for publication.

\bibliography{reference}
\bibliographystyle{acl_natbib}

\out{
\noindent LaTeX formats citations and references automatically using the bibliography records in your .bib file, which you can edit via the project menu. Use the cite command for an inline citation, e.g.  \cite{Hao:gidmaps:2014}.

For data citations of datasets uploaded to e.g. \emph{figshare}, please use the \verb|howpublished| option in the bib entry to specify the platform and the link, as in the \verb|Hao:gidmaps:2014| example in the sample bibliography file.
}


\section*{Author contributions statement}
S.F., S.H., and T.M. conceived the experiment(s), K.T., S.H., and T.M. conducted the experiment(s), and K.T., S.H., E.A., and S.W. analyzed the results. E.A., S.Y., S.W., and S.F. contributed to the study design and wrote the main manuscript text. All authors reviewed the manuscript. 

\section*{Additional information}
The corresponding author has no competing interests.



\out{
\begin{figure}[ht]
\centering
\includegraphics[width=\linewidth]{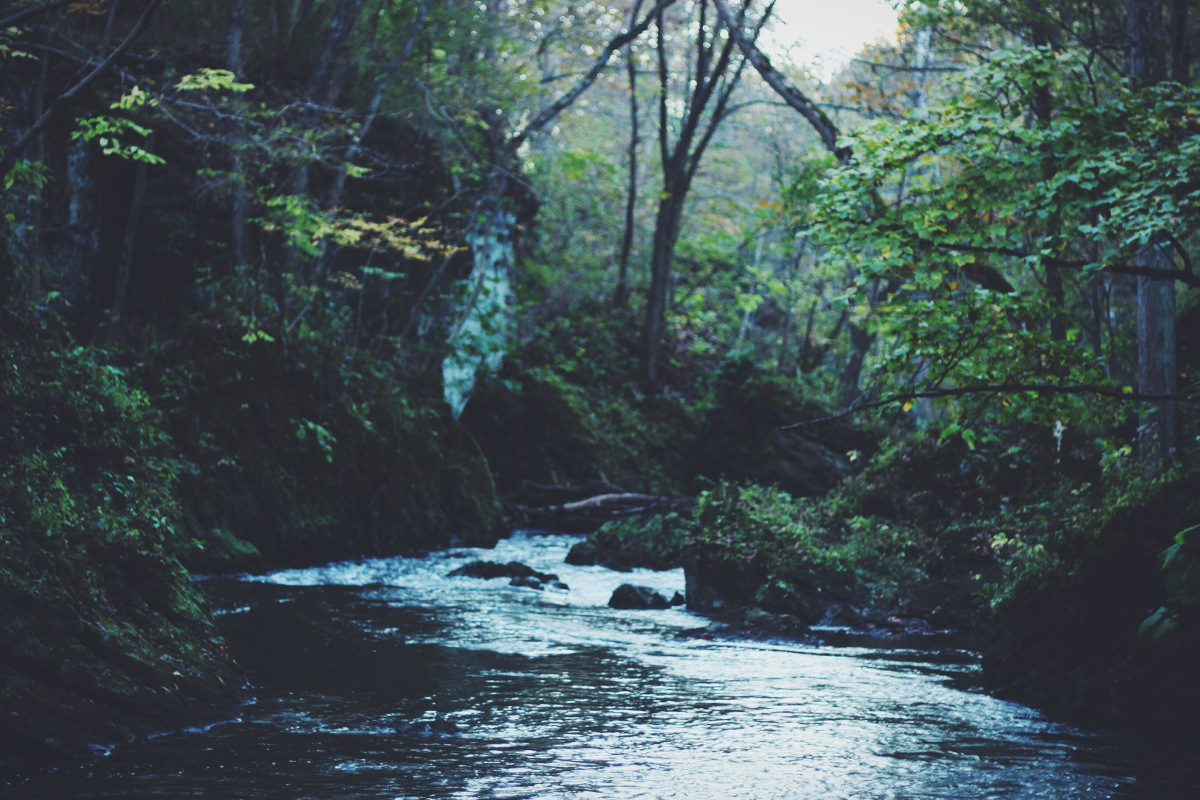}
\caption{Legend (350 words max). Example legend text.}
\label{fig:stream}
\end{figure}

\begin{table}[ht]
\centering
\begin{tabular}{|l|l|l|}
\hline
Condition & n & p \\
\hline
A & 5 & 0.1 \\
\hline
B & 10 & 0.01 \\
\hline
\end{tabular}
\caption{\label{tab:example}Legend (350 words max). Example legend text.}
\end{table}

Figures and tables can be referenced in LaTeX using the ref command, e.g. Figure \ref{fig:stream} and Table \ref{tab:example}.
}

\end{document}